\documentclass{article}
\usepackage{spconf,amsmath,graphicx}
\usepackage{booktabs}
\usepackage{algorithm, algorithmic}
\usepackage{enumitem}
\usepackage[export]{adjustbox}
\usepackage{hyperref}
\title{LSTM-based Video Quality Prediction Accounting for Temporal Distortions in Videoconferencing Calls}
\name{Author(s) Name(s)}

\name{Gabriel Mittag, Babak Naderi, Vishak Gopal, Ross Cutler}

\address{Microsoft Corp.}

\begin{document}
%
\maketitle
\begin{abstract}
Current state-of-the-art video quality models, such as VMAF, give excellent prediction results by comparing the degraded video with its reference video. However, they do not consider temporal distortions (e.g., frame freezes or skips) that occur during videoconferencing calls. In this paper, we present a data-driven approach for modeling such distortions automatically by training an LSTM with subjective quality ratings labeled via crowdsourcing. The videos were collected from live videoconferencing calls in 83 different network conditions. We applied QR codes as markers on the source videos to create aligned references and compute temporal features based on the alignment vectors. Using these features together with VMAF core features, our proposed model achieves a PCC of 0.99 on the validation set. Furthermore, our model outputs per-frame quality that gives detailed insight into the cause of video quality impairments. The VCM model and dataset are open-sourced at \url{https://github.com/microsoft/Video_Call_MOS}.
\end{abstract}
\section{Introduction}
One of the most important performance metrics for evaluating videoconferencing (VC) system is the perceived video quality of a call. The ground truth quality can only be assessed in subjective experiments, for example, according to ITU-T (International Telecommunication Union) Recommendation P.910 \cite{p910}, in which participants are asked to view sample videos and rate their quality. The average rating across all participants gives the so-called Mean Opinion Score (MOS). Recently, crowdsourcing has become increasingly popular for conducting video-quality experiments. Implementations, such as P.910 Crowd \cite{naderi2022}, allow for conducting accurate video-quality experiments with high reproducibility.

Because subjective experiments are time-consuming and costly, signal-based prediction models have been developed that are able to predict the quality of a given video. The most commonly used video quality metrics are PSNR (Peak Signal to Noise Ratio), SSIM \cite{ssim}, MS-SSIM \cite{msssim}, and VMAF \cite{vmaf}, which are computed on a frame basis and with subsequent averaging across all frame scores to give the overall quality. Although other pooling mechanisms have been studied in order to take recency effects into account, in most studies mean pooling achieved the best performance \cite{pool, vmaf2}. VMAF is the only of those metrics that include a temporal component, which considers temporal masking effects. However, video transmitted during VC calls can be affected by a number of temporal distortions that are perceived as frame freezes, frame skips, frame rate variations (e.g., video is played back faster after delayed packets arrive), or generally low frame rate. These kinds of distortions cannot be covered by the aforementioned quality metrics as they are mostly frame-based and do not take time dependencies into account. 

More recently, deep learning-based video quality models such as VSFA \cite{VSFA}, DeepVQA \cite{deepvqa}, CompressedVQA \cite{compressedvqa}, and C3DVQA \cite{C3DVQA} have been introduced. Although some of them take temporal-memory effects or temporal masking into account and apply more advanced pooling mechanisms, they are neither designed for nor trained on videos with temporal distortions as they appear in VC calls. In general, most of the temporal video quality modeling has been focused around streaming applications \cite{temporal_effects} that are subject to different degradation types than VC systems. The MOVIE \cite{movie}, ST-MAD  \cite{stmad}, and STRRED \cite{strred} models consider visual perception of motion artifacts. VQM\_VFD \cite{VQM_VFD} takes the impact of variable frame delays (including freezes) into account by combining 8 different parameters with a shallow neural network. ITU-T Rec. P.1203 \cite{Robitza2018HTTPAS} is a parametric bitstream-based audiovisual quality prediction model for adaptive streaming services. ATLAS \cite{atlas} includes rebuffering-aware and memory-driven effects. The model in \cite{recurrent} also considers rebuffering in video streams by applying recurrent and dynamic neural networks. ST-GREED \cite{stgreed} applies entropy variations between reference and distorted videos to consider quality variations caused by frame rate changes. 

Another challenge for full reference VC quality prediction is the alignment of the received video with the original video. In the VQM\_VFD model, a heuristic algorithm based on pixel-by-pixel comparisons is applied, which is prone to alignment errors. Instead, we apply markers to the source videos, which allows for reading the corresponding reference frame index from the degraded frames. To the best of our knowledge, this is the first work in which the alignment vector is leveraged to model the impact of temporal distortions as they occur during video calls with a data-driven approach. To this end, we use a recurrent neural network and directly input features based on the alignment vector as a sequence, fused together with image quality metrics.

\section{Dataset}
\label{sec:dataset}
To train and evaluate the presented model, a new dataset with recordings from live Microsoft Teams calls was created. The recorded videos are based on ten source videos. Eight of the videos contain one person speaking into the camera and two of the videos contained two people talking to each other. The video resolution of the source videos is 1920x1080 with a frame rate of 30 FPS. In a preprocessing step, two QR codes, containing the original frame index, were drawn onto the source videos to make an alignment between degraded and reference videos possible (see Subsection~\ref{sec:align}).

To create the dataset, calls between two machines and various emulated real-call network scenarios were conducted. The scenarios comprised of different fixed bandwidths, burst loss, random loss, fluctuating bandwidth, cross traffic, and combinations of these conditions. During the call, the video bitrate and resolution are adapted depending on the network condition. As a consequence, the quality of the received videos changes over time and may contain different bitrates, switches between video resolutions, frame rate variation, and frozen or skipped frames. The degraded videos were captured on the receiving side with 30 FPS and cropped from the call duration of 2 minutes to clips of 6 seconds, where a random segment of the call was chosen as the clip. 
%
%
In that way, 1628 videos with 83 different network profiles were collected. The videos were annotated for their overall quality on an ACR (Absolute Category Rating) scale according to P.910 \cite{p910} via crowdsourcing using the P.910 Toolkit\footnote{https://github.com/microsoft/P.910} \cite{naderi2022}. 
On average, 17 valid votes were collected per clip. To split the dataset, the videos that are based on two of the ten source videos were selected for validation, resulting in 1299 training set videos and 329 validation set videos, where the training and validation sets are sharing the same network profiles but having different source videos.
%
%
\section{Method}
\subsection{Alignment}
\label{sec:align}
Most video quality models compare reference video frames with degraded frames, but this is challenging for live video calls as recordings are not aligned with the original reference video. We added QR codes to the source videos to enable alignment, but reading codes in poor network conditions can be difficult. To improve detection, we added two identical QR codes to each video, one in the top-left and one in the bottom-right corner.

To create an aligned reference video, the QR codes from the degraded videos are read for each frame. This gives an index vector $r(i)$ with the reference frame indices for each degraded frame. In the case of frozen frames the vector contains consecutive repetitions of the same reference frame index, in the case of frame skips the vector is missing certain reference frame indices. Then we create a new reference video with the frame order given by $r(i)$. 
\subsection{Features}
Because VMAF is widely used and has been shown to give robust quality predictions for videos, we use the same input features as VMAF to cover the image quality of the individual frames. However, these features cannot cover temporal quality degradation caused by freezes/skips. For this reason, we add two additional frame freeze features that are based on the reference frame indices $r(i)$.
\paragraph*{VMAF core features}
\begin{itemize}[nosep]
    \item VIF \cite{Sheikh2004ImageIA} -- Image quality metric that measures the information of fidelity loss between degraded and reference images. VMAF uses a modified version where the loss of fidelity in each scale is included as an elementary metric \cite{vmaf}.
    \item ADM \cite{Li2011ImageQA} -- Image quality metric that measures the loss of details that affect the content visibility and additionally the redundant impairment that distracts the viewer's attention.
    \item Motion2 \cite{vmaf} -- Temporal metric that measures the temporal difference between adjacent frames by calculating the average absolute pixel difference for the luminance component.
\end{itemize}
\paragraph*{Frame freeze feature}
\begin{itemize}[nosep]
    \item Frame skips -- Measures the reference frame indices difference between consecutive frames as follows: $s_i = r_i-r_{i-1}$ for $i \in [1,N-1]$ with $s_0=0$, where $r(i)$ is the reference indices vector and $N$ is the number of frames of the degraded video.
    \item Frame freezes -- Simple measure to capture the number of consecutive frozen frames according to Algorithm \ref{alg:algorithm}.
    
    
\end{itemize}

\begin{algorithm}[ht!]
\caption{Pseudo code for computing freeze feature}
\label{alg:algorithm}
\textbf{Input}: Reference frame index $r_i$ for each degraded frame index $i\in [0,N-1]$, where $N$ is the number of frames of the degraded video. \\
\textbf{Output}: Freeze value $f_i$.\par
\begin{algorithmic}[1] 
\STATE $f_{0}=0$ 
\FOR{$i = 1$;\ $i < N$;\ $i=i+1$}
\IF{$r_i==r_{i-1}$}
\STATE $f_i=f_{i-1}+1$
\ELSE
\STATE $f_i=0$
\ENDIF
\ENDFOR
\end{algorithmic}
\end{algorithm}

\subsection{Model}
\begin{figure}[!ht]
    \begin{center}
        \includegraphics[width=0.75\linewidth, right]{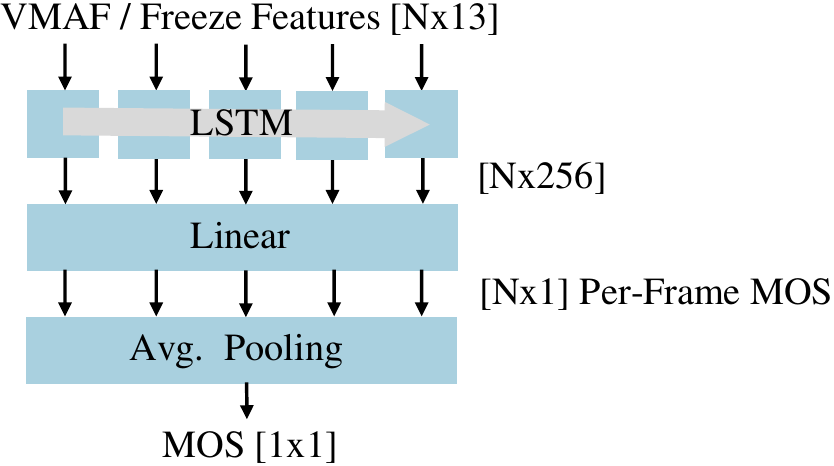}
    \end{center}
    \vspace*{-5mm}
    \caption{Block diagram of the proposed neural network.}
    \label{fig:model}
\end{figure}

As a model, we apply a simple LSTM (Long Short-Term Memory) network, as it is well-suited for modeling sequences and time dependencies. We use 6 layers with each 256 hidden units, although in our experiments smaller models (e.g., with 1 or 2 layers and 128 hidden units) reached similar performances. The model architecture is shown in Figure~\ref{fig:model}. Besides predicting the overall video quality, another goal of this work is to output the quality per frame, including the impact of temporal effects, such as frozen frames. In order to force the model to output a single quality value per frame, instead of applying time pooling first and then using a linear layer for the final output, we apply the linear layer directly after the LSTM to reduce the feature size from 256 to 1. Then we apply an average pooling layer to compute the final overall video quality. In this way, also the common problem of choosing a suitable time pooling method is solved, as the LSTM learns to weigh the impact of the individual frames on the overall quality automatically.
%
\section{Results and Discussion}
In this section, the results of the model on the validation set are presented. At first, we present the results compared to the default VMAF model. Then, we present the results of an ablation study, where we compare the results for different input features and a retrained VMAF model. All of the presented scores were mapped to the subjective MOS with a Sigmoid function. The mapped values were then used for the plots and computing the figures of metrics PCC (Pearson's correlation coefficient) and RMSE (root-mean-square error). The mapping is  particularly necessary for VMAF, SSIM, and PSNR. For example, the scores of the default VMAF model range between 0--100, while the videos in our dataset are rated in MOS between 1--5. All of the presented results are computed on a per-file base. 
\subsection{Validation set results}
\begin{figure}[!ht]
\begin{minipage}[b]{.48\linewidth}
  \centering
  \centerline{\includegraphics[width=4.0cm]{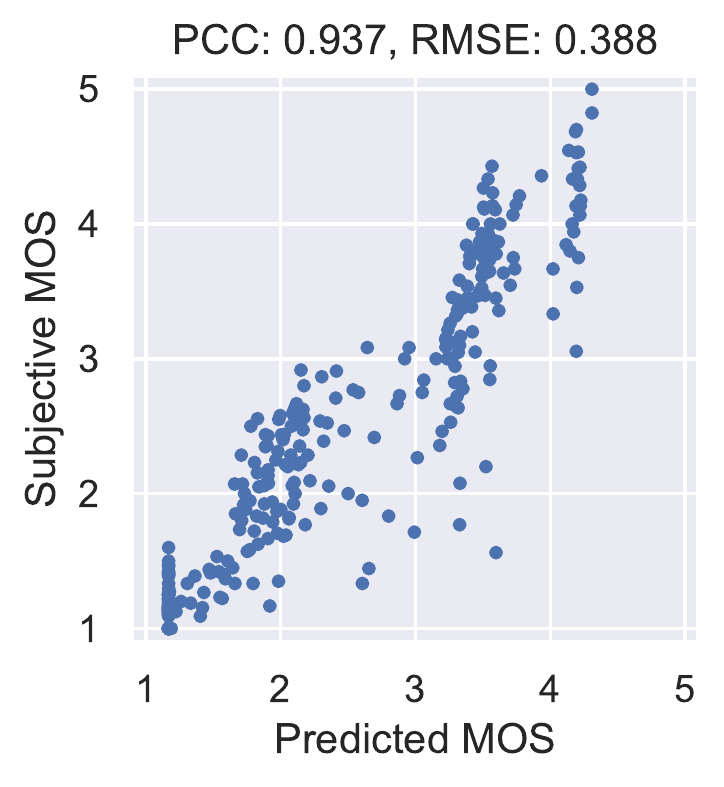}}
  \centerline{\small(a) VMAF}\medskip
\end{minipage}
\hfill
\begin{minipage}[b]{0.48\linewidth}
  \centering
  \centerline{\includegraphics[width=4.0cm]{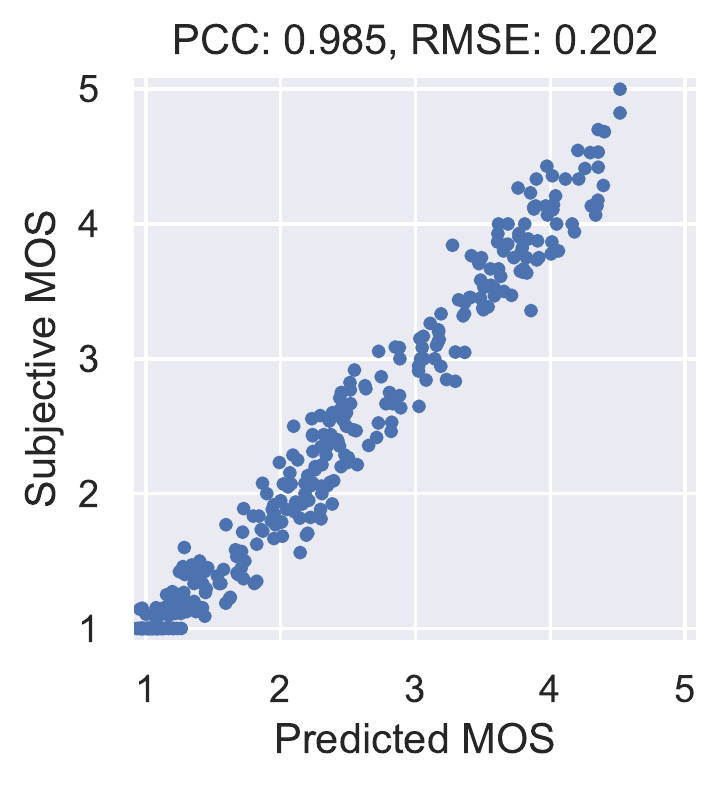}}
  \centerline{\small(b) Proposed} \medskip
\end{minipage}
\vspace*{-5mm}
\caption{Scatterplot showing the ground truth and predicted MOS values per file on the validation set.}
\label{fig:reslts}
\end{figure}
Figure~\ref{fig:reslts} shows the scatter plots of the predicted values for the default VMAF model \textit{(a)} and the proposed LSTM with VMAF and temporal features as input \textit{(b)}. The proposed model outperforms the default VMAF model with a PCC increase from 0.94 to 0.99. The plots also show that most of the VMAF outliers are caused by overestimating the video quality, presumably because these video samples contain video freezes that are not captured by VMAF.
\begin{figure}[!ht]
\begin{minipage}[h]{1.0\linewidth}
\centering
\includegraphics[width=0.95\linewidth]{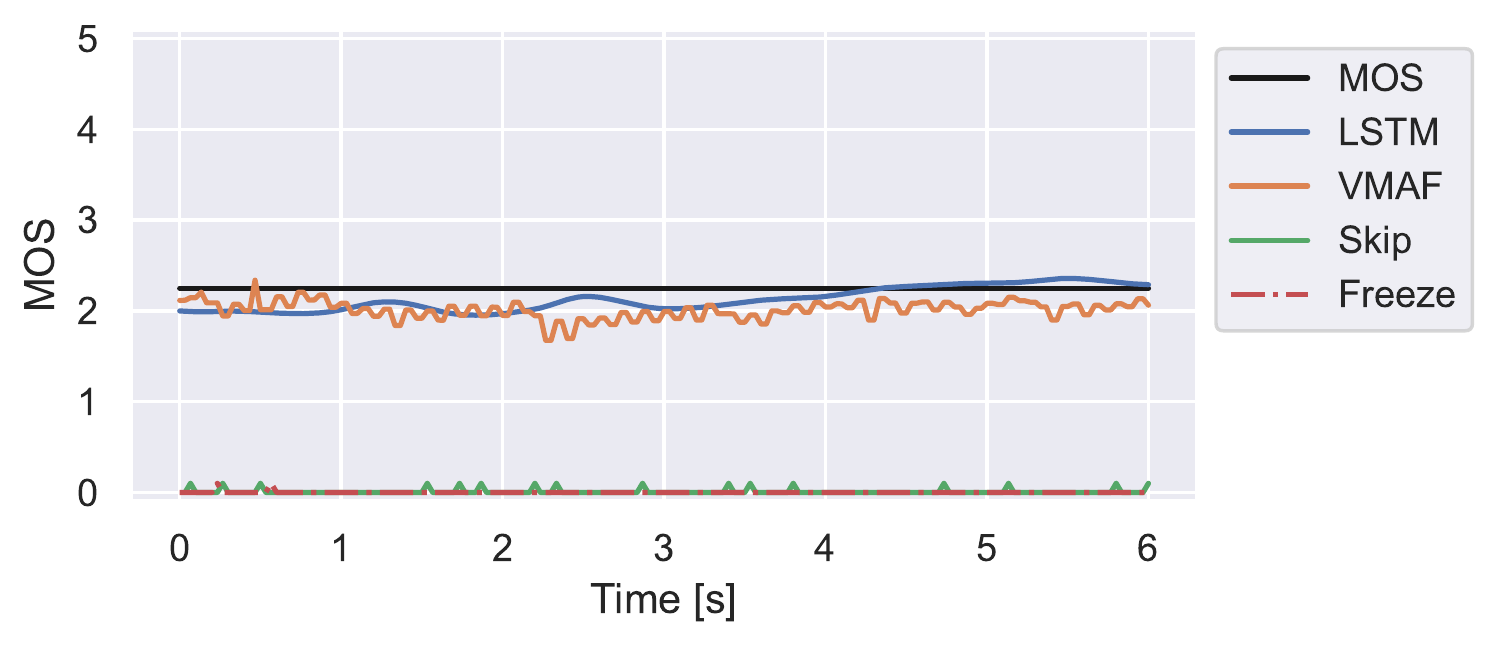}
\end{minipage}
%
\begin{minipage}[h]{1\linewidth}
\centering
\includegraphics[width=0.95\linewidth]{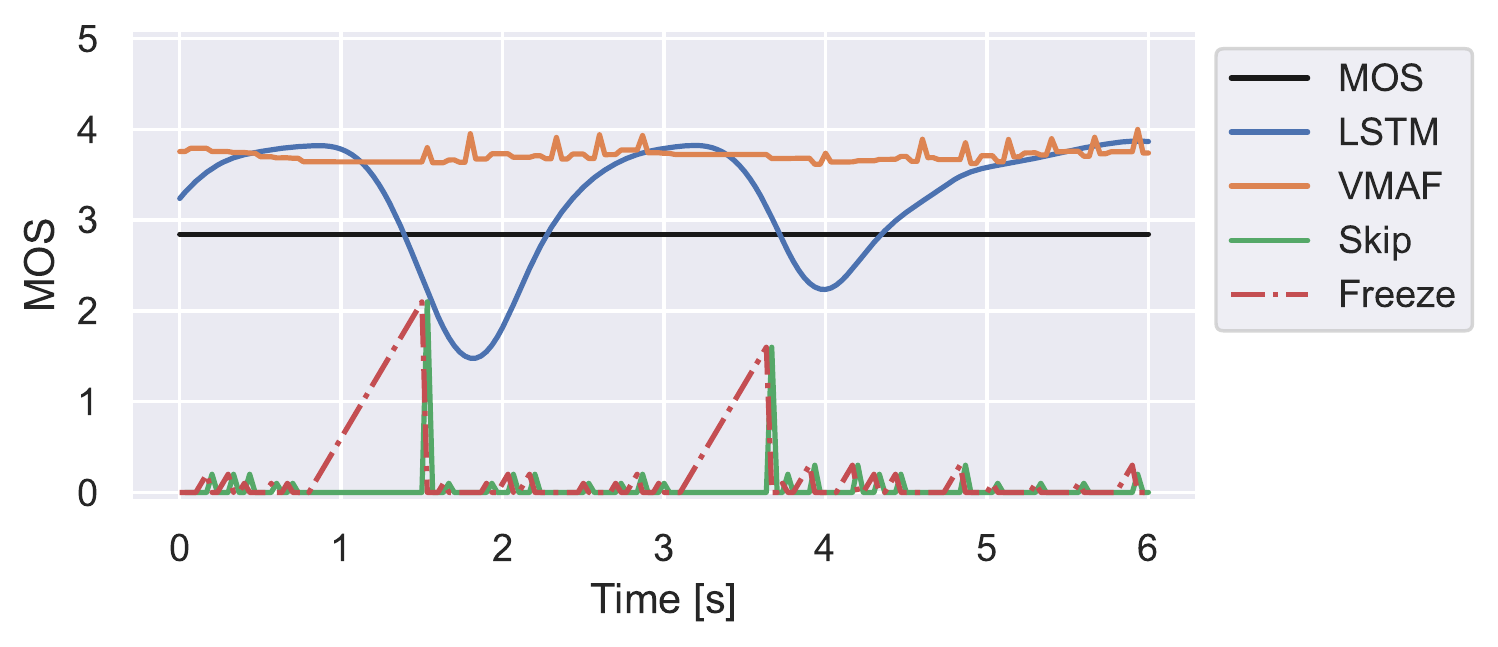}
\end{minipage}   
%
%
\vspace*{-5mm}
\caption{Per-frame quality over time for two example videos.}
\label{fig:examples}
\end{figure}
To further analyze this presumption, we plot the per-frame video quality over time for two example videos in Figure~\ref{fig:examples}. The black line represents the overall MOS rating from the crowdsourcing experiment, the orange line shows the per-frame VMAF score, and the blue line shows the per-frame quality of the proposed LSTM model. In addition to the quality predictions, we also plot the frame freeze and skip features, which are shown in frames divided by 10 (i.e., in the plot a ``freeze'' value of 1 represents 10 consecutive frozen frames which corresponds to 0.33 seconds). In the top graph, a video without any major freezes or frame skips is shown. Both VMAF and the LSTM predict the quality to be close to the ground truth MOS. The bottom graph shows a video with two video freezes. VMAF clearly overestimates the quality of this video. The LSTM begins with a similar prediction as VMAF but lowers the quality prediction once the first frame freeze starts. After the freeze has ended and the video continues playing, the LSTM increases its quality prediction again. By considering the freezes the model reaches an average video quality prediction close to the ground truth MOS. These results show that the per-frame quality prediction of the proposed model works very well and can be used to analyze the root cause of videos with poor quality.
\subsection{Ablation study}
In this subsection, we study the influence of the different input features on the results and compare the model to a retrained VMAF model. For the retraining of VMAF, we used the default SVR (Support Vector Regression) settings with $\gamma = 0.05$, $C = 4.0$, and $\nu = 0.9$ and the 11 VMAF core features VIF (vif\_scale0, vif\_scale1, vif\_scale2, vif\_scale3), ADM (adm2, adm\_scale0, adm\_scale1, adm\_scale2, adm\_scale3), and Motion (motion, motion2). Table~\ref{tab:results} shows the validation set results for the default and retrained VMAF. We also show the results for the image quality metrics PSNR, SSIM, and MS-SSIM, which are commonly used as baseline metrics. It should be noted that the results for the LSTM show the average PCC and RMSE over three training runs for each feature configuration, where the best checkpoint of each training run was selected. 
\begin{table}[htb]
\caption{Validation set results for baseline and proposed model using different input features.}
\vspace{0.2cm}
\label{tab:results}
\centering
\small
\begin{tabular}{@{}l|c|c@{}}
\toprule
\textbf{Model}                                           & \textbf{PCC}          & \textbf{RMSE}          \\ \midrule
PSNR                                                     & 0.928                 & 0.415                  \\
SSIM                                                     & 0.951                 & 0.342                  \\
MS-SSIM                                                  & 0.946                 & 0.360                  \\
VMAF                                                     & 0.937                 & 0.388                  \\
VMAF (Retrained)                                         & 0.965                 & 0.289                  \\
LSTM - VIF / ADM                                         & 0.956                 & 0.335                  \\
LSTM - VIF / ADM / Motion                                & 0.983                 & 0.218                  \\
LSTM - VIF / ADM / Skip / Freeze                         & 0.984                 & 0.213                  \\
LSTM - VIF / ADM / Skip                                  & 0.979                 & 0.253                  \\
LSTM - VIF / ADM / Freeze                                & 0.983                 & 0.213                  \\
LSTM - VIF / ADM / Motion / Skip / Freeze                & \textbf{0.985}        & \textbf{0.206}         \\
LSTM - '' / '' / '' / '' / '' / PSNR / SSIM              & 0.983                 & 0.211                  \\ \bottomrule
\end{tabular}
\end{table}
Interestingly, SSIM outperforms MS-SSIM and the default VMAF model with a PCC of 0.951. Retraining VMAF notably improves the performance and increases the PCC from 0.937 to 0.965. The LSTM with the same VMAF core input features (VIF/ADM/Motion) outperforms the retrained VMAF model with a PCC of 0.983, showing that LSTM models are better suited to model the temporal distortions in VC calls than the default SVR used by VMAF. It can further be noted that adding either of the temporal features (Motion, Skip, or Freeze) increases the results from a PCC of 0.956, when only the image quality metrics VIF and ADM are applied, to a PCC of around 0.98.

The LSTM seems to be able to consider frame freezes or skips with any of the temporal features. The purpose of the VMAF Motion feature is to capture the temporal masking of coding artifacts and is calculated solely on the basis of the reference video. In the case of high motion values, VMAF will judge distortions less severely as the human eye does not perceive some distortions during high motion between frames. However, since we aligned our reference videos to the degraded video, the reference also contains the frame freezes of the degraded video and thus leads to a value of zero for frozen frames as the frames are exactly identical. Therefore, the LSTM can use the motion feature to detect video freezes instead of only using it to capture the temporal masking effect. That being said, the Motion feature is not able to capture frame rate variations. However, in practice, such changes are usually preceded by frame freezes which already impact the perceived quality. Therefore no significant performance improvements could be observed on our dataset when the skip or freeze features were added. Furthermore, the LSTM could confuse a video with still sequences with frozen frames when Motion is used as the only temporal feature (in cases where the reference alignment is not done via video markers as they will lead to non-zero Motion values). Although the performance difference between the temporal features is only marginal, overall, the best performance can be achieved by including all of them, in particular when considering the RMSE. Adding the Skip and Freeze features decreases the RMSE from 0.218 to 0.206, compared to when only Motion is used. Adding PSNR and SSIM to the LSTM model did not further improve the results.
\section{Conclusions}
We proposed the open-sourced VCM model based on a simple but highly efficient data-driven approach for modeling temporal distortions in VC videos. An open-source dataset with live video calls in 83 different network conditions was created in which we preprocessed the source videos with QR code markers for the alignment between degraded and reference videos. We show that by using VMAF core features to cover the non-temporal visual quality of the video frames together with temporal features based on the alignment vector, the LSTM learns to model temporal distortions automatically and achieves very high accuracy with a PCC of 0.99 on the validation set. Our model further outputs per-frame MOS that gives detailed insight into the cause of degraded quality. In future work, the VMAF core features could be replaced with an end-to-end deep learning approach to further improve the results.
%
\bibliographystyle{IEEEbib}
\bibliography{strings,refs}

\end{document}